\documentclass[preprint,showpacs,preprintnumbers,amsmath,amssymb,showkeys]{revtex4}
\usepackage{graphicx}
\usepackage{dcolumn}
\usepackage{bm}

\begin{document}
\renewcommand{\thesection}{\arabic{section}}

\bibliographystyle{apsrev}

\title{\center{ Elliptical torii in a constant magnetic field\\
}}

\author{M. Encinosa }
\email{encinosa@cennas.nhmfl.gov}
\author{M. Jack}%

\affiliation{ Florida A\&M University\\Department of Physics \\
Tallahassee FL 32307}
\newpage


\begin{abstract}
The Schrodinger equation for an electron on the surface of an
elliptical torus in the presence of a constant azimuthally
symmetric magnetic field is developed. The single particle
spectrum and eigenfunctions as a function of magnetic flux through
the torus are determined and it is shown that inclusion of the
geometric potential is necessary to recover the limiting cases of
vertical strip and flat ring structures.
\end{abstract}

\maketitle

\section{\label{sec:level1}Introduction}

Toroidal nanostructures  present intriguing possibilites for use
as nano-device elements \cite{shea,sasaki,sano,zhang}. In addition
to Aharanov-Bohm and persistent azimuthal current effects known to
exist in quantum rings {\cite{filikhin,
gylfad,viefers,latil,sasaki2,gridin,pershin}, toroidal structures
allow for motions around the minor radius of the torus subject to
boundary conditions distinct from those for flat rings (Figs.
1-3).

For hollow torii, electrons are thought to be localized near the
surface of the object. The restriction to motion near a surface
has interesting manifestations. Recent work \cite{encmott,
encmott2} on a torus of major radius $R$ and circular cross
section of radius $a$, has shown that a surface dependent
geometric potential $V_C$
\cite{jenskoppe,dacosta1,dacosta2,exnerseba,matsutani,burgsjens}
is important even for electrons that can wander substantial
distances away from the surface and should be employed as an
effective potential when considering two dimensional problems on
curved surfaces. Curvature effects are more pronounced for
elliptical torii ($ET^2$) which can behave in different ways than
a torus with  circular cross section.  A recent paper by Gravesen,
Willatzen and Lew Yan Voon in this journal \cite{gravesen} dealt
with particles constrained to motion on surfaces of revolution,
including elliptical torii. There the influence of toroidal
eccentricity and $V_C$  were shown to strongly affect system
eigenvalues and eigenfunctions. This work is concerned with the
extension of \cite{gravesen} to include a constant magnetic field
along the z-axis, and the role $V_C$ plays in recovering the
ribbon (taken to be an infinitely thin vertical strip of radius
$R$ and height $2b$) and flat ring (an annular region of inner
radius $R-a$ and outer radius $R+a$) limits.

This paper is organized as follows: section 2  presents the
geometry leading to, and the  formalism by which, the Hamiltonian
on $ET^2$ is derived inclusive of a magnetic field ${\mathbf B} =
B_0 {\mathbf e}_z$. The methodology presented in section 2 differs
from that given in \cite{gravesen} because here the magnetic field
is to be incorporated via the minimal prescription  so it proves
advantageous to first derive the gradient operator rather than
directly employ
\begin{equation}
\nabla^2= g^{-{1 \over 2}}{\partial \over \partial q^i} \bigg [
g^{1 \over 2}\ g^{ij}{\partial \over \partial q^j} \bigg ].
\end{equation}
 In section 3 numerical results are given as curves of single particle
ground state energies $\varepsilon_0$ as functions of magnetic
flux $\gamma$ for elliptical torii of several eccentricities.
Section 4 is reserved for conclusions and suggestions for future
work.

\section{Development of the Hamiltonian on $ET^2$}
As noted above, including a vector potential $\mathbf A$ in the
Hamiltonian makes it advantageous to derive the gradient operator
and proceed to include  $\mathbf A$ by the minimal prescription.

Let (${\mathbf e}_{\rho}, {\mathbf e}_{\phi},{\mathbf e}_{z}$) be
cylindrical coordinate system unit vectors. Parameterize points
near an elliptical toroidal surface of major radius $R$ and minor
radii
 $a$ and $b$ by
\begin{equation}
 {\mathbf x}(\theta,\phi,q)=(R + a \ {\rm cos}  \theta ){\mathbf e}_{\rho} +b\  {\rm sin}
\theta{\mathbf e}_z + q{\mathbf e}_{n}
\end{equation}
with  ${\mathbf e}_{n}$ the unit normal to $ET^2$ to be defined
momentarily and $q$ the coordinate  measuring the distance from
the surface. In what follows the major radius $R$ will be set to
$R = 500 \AA$, which is in accordance with fabricated structures
\cite{lorke,garsia,zhang}. From Eq. (2)
\begin{equation}
d{\bf x}= P d\theta {\mathbf e}_1+W d\phi {\mathbf e}_{\phi}
+dq{\mathbf e}_n +qd{\mathbf e}_n
\end{equation}
with
\begin{equation}
P=(a^2 {\rm sin^2} \theta + b^2 {\rm cos^2} \theta)^{1/2}
\end{equation}
\begin{equation}
W=1+ \alpha \ {\rm cos } \theta
\end{equation}
\begin{equation}
{\mathbf e}_1 = {1 \over P}(-a \ {\rm sin\theta}{\mathbf e}_{\rho}
+ b \ {\rm cos\theta}{\mathbf e}_z)
\end{equation}
and
\begin{equation}
{\mathbf e}_n = {1 \over P}(a \ {\rm sin\theta}{\mathbf e}_z + b \
{\rm cos\theta}{\mathbf e}_{\rho}).
\end{equation}
The unit vectors (${\mathbf e}_1, {\mathbf e}_n$) are tangent to
$ET^2$ in the direction of increasing $\theta$ and $\phi$
respectively.

 From
\begin{equation}
d{\mathbf e}_n = {ab \over P^2} \ {\mathbf e}_1 d\theta + {b \over
P} \ {\rm cos\theta}{\mathbf e}_{\phi}d\phi,
\end{equation}
the vector differential line element can be found from Eqs. (8)
and (3), leading to a gradient operator
\begin{equation}
\nabla = {1 \over P(1+k_\theta q)} {\mathbf e}_1 {\partial \over
\partial \theta} +{1 \over W(1+k_\phi q)} {\mathbf e}_\phi {\partial \over \partial \phi}+
{\mathbf e}_q {\partial \over
\partial q}
\end{equation}
with $k_\theta, k_\phi$ the principle curvatures given by
\begin{equation}
k_\theta = {ab \over P^3}, \ \  k_\phi= {b \ {\rm cos}\theta \over
WP}.
\end{equation}

 The vector potential (working in the Coulomb $\nabla \cdot
{\mathbf A}=0$ gauge) appropriate to $B_0 {\mathbf e}_z$ is
\begin{equation}
{\mathbf A} = {B_0 \over 2} \bigg( W + {qb \ {\rm cos} \theta
\over P} \bigg ){\mathbf e}_\phi.
\end{equation}
The Schrodinger equation ($\hbar = m = 1$)
\begin{equation}
{1\over 2}\bigg({1\over i} \nabla + q \bm A \bigg)^2 \Psi = E\Psi
\end{equation} which results from Eqs. (9), (11) and (12) can be
reduced to  a concise and dimensionless form by first
\begin{list}{}{\itemsep -6pt}
\item a. noting
\begin{equation}
{\partial {\mathbf e}_1 \over \partial \theta}= -{ab\over
P^2}{\mathbf e}_n
\end{equation}
\begin{equation}
{\partial {\mathbf e}_n \over \partial \theta}= +{ab\over
P^2}{\mathbf e}_1
\end{equation}
\begin{equation}
{\partial {\mathbf e}_1 \over \partial \phi}= -{a \  {\rm
sin}\theta\over P}{\mathbf e}_\phi
\end{equation}
\begin{equation}
{\partial {\mathbf e}_n \over \partial \phi}= +{ b \ {\rm
cos}\theta \over P }{\mathbf e}_\phi,
\end{equation}
\vskip 8pt \item b. setting $\alpha = {a/R}$, $\beta = {b/R}$,
$\varepsilon = 2Ea^2$, \item c. letting $D(\theta) = {P(\theta)/
R}$, $p(\theta) = P(\theta)/a$ and $F(\theta) = 1 + \alpha \ \rm
cos\theta$, \item d. defining $\gamma = .263 B_0$, with $B_0$ in
Teslas, which is the conversion factor for an $R = 500 \AA$ torus,
\item e. performing the well-known procedure for obtaining $V_C$
(which will appear below as the scaled dimensionless function
$U_C$) for which the reader is directed to the relevant references
\cite{jenskoppe,dacosta1,dacosta2,exnerseba,matsutani,burgsjens,
duclosexner,bindscatt,goldjaffe,ouyang,popov,midgwang,clarbrac,schujaff},
 \item f. noting that the azimuthal symmetry of the
problem allows for the eigenfunction on $ET^2$ to be taken as
$\Psi(\theta,\phi) = \psi(\theta)exp \ [i\nu\phi]$.
\end{list}
Applying the conventions and procedures listed above results in a
pair of equations for the surface and normal variables in the $q
\rightarrow 0$ limit
 \begin{widetext}
\begin{equation}
{\partial^2 \psi \over \partial \theta^2} - \bigg ( {\alpha \ {
\rm sin}\ \theta \over F(\theta)}+{{\alpha^2-\beta^2}\over
 D^2(\theta)} \bigg ) {\partial \psi \over \partial\theta} -
 \bigg [{D^2(\theta) \nu^2 \over F^2(\theta)}+ U_C(\theta)
+{\gamma^2F^2(\theta)\alpha^2 p^2(\theta) \over 4}+ \gamma \nu
\alpha^2  p(\theta) -\varepsilon \bigg ] \psi =0,
\end{equation}
\end{widetext}

\begin{equation}
-{1 \over 2}{\partial^2  \chi_n \over \partial q^2} + V_n(q)
\chi_n =E_n \chi_n.
\end{equation}

 The scaled dimensionless curvature potential $U_C$ appearing in Eq. (17) is
\begin{equation}
U_C(\theta) = -{1\over 4}\bigg( {\alpha^2\beta^2 \over
D^4(\theta)} + {\beta^2 {\rm cos^2}\theta  \over F^2(\theta)}\bigg
) .
\end{equation}
The normal confining potential $V_n(q)$ can be chosen to take any
convenient form, and it is apparent in the $q \rightarrow 0$ limit
that the  surface Schrodinger equation is independent of the
choice. The independence of surface observables on $V_n(q)$ has
also been shown to be a good approximation when the particle is
allowed to move in a finite thin layer by a basis set calculation
in the full three dimensional space \cite{encmott2}. The results
that follow will not include $E_n$; it was also shown in
\cite{encmott2} there is negligible state mixing in the $q$ degree
of freedom so that the surface spectrum is essentially independent
of $E_n$.

\section{Numerical results}
The Schrodinger equation given by Eq. (17) is invariant under
$\theta \rightarrow -\theta$ making it possible to separate its
eigenfunctions into even and odd $\theta$-parity states by proper
choice of initial conditions at $\theta = 0$ \cite{fpl}.
Numerically it proves convenient to generate two linearly
independent solutions of Eq. (17) with initial conditions
\begin{equation}
\psi(0) = 1, \ \psi'(0)= 0 \ \rightarrow \psi_A
\end{equation}
\begin{equation}
\psi(0) = 0, \ \psi'(0)= 1 \ \rightarrow \psi_B
\end{equation}
and insisting upon
 \begin{equation} A\psi_A(0)+B\psi_B(0) =
A\psi_A(2\pi)+B\psi_B(2\pi)
\end{equation}
 \begin{equation} A\psi'_A(0)+B\psi'_B(0) =
A\psi'_A(2\pi)+B\psi'_B(2\pi).
\end{equation}
Eqs. (22) and (23) can be rearranged into a homogeneous linear
system for the coefficients $A,B$  and the entire spectrum quickly
determined by a simple loop over $\varepsilon$ appearing in Eq.
(17).

 Single particle ground state  energy $\varepsilon_0$ plots as a
function of $\gamma$ are given for three toroidal eccentricities
in Figs. (4-6). The most interesting  feature  emerging from
inspection of Figs. (4) and (5) is the necessity of including
$U_C$ to approach the  ring and ribbon limit. However, as shown in
Figs. (7) and (8), the ring and ribbon limits do differ in detail
when compared to the curves for the elliptical torii; because
persistent currents are sensitive to the shape of the free energy
as a function of flux, the differences in detail may prove
important upon a more extensive calculation of spectra. Comparison
of the curves shown in Fig. (6) with the values of the limiting
cases appearing in the previous two figures indicate clearly that
a torus with circular cross-section cannot realistically be
approximated with a ring or ribbon, and that the $U_C$ alters the
structure of $\varepsilon_0(\gamma)$ substantially.

\section{Conclusions}
In this work single-particle ground state energies $\varepsilon_0$
as a function of magnetic flux $\gamma$ were calculated for three
elliptical torii of very different character. The curves indicate
that any attempt to model toroidal structures by approximating
them as two-dimensional  ribbons or rings must include the
geometric potential in order to recover the respective two
dimensional limits. This result is  surprising in that
 $V_C$ is large only near the regions of substantial
 curvature and is negligible over most of the structure.
  Nevertheless, it has considerable effect on $ET^2$
eigenvalues, and its omission leads to disagreement with the flat
limits. The inclusion of $V_C$, however, does not simply trivially
reproduce the flat limits for the toroidal eccentricities
investigated here; the $\varepsilon_0 (\gamma)$ curves show
differing peak heights and locations than those of the limiting
cases.

An interesting natural extension of this work is the addition of
an off-axis component of the applied magnetic field. Those cases
are perhaps best suited to a basis set method as that employed in
\cite{T2mag} subject to modification of the integration measure
arising from toroidal eccentricity. The interplay of the magnetic
field with regions of substantial curvature may yield interesting
mixing of azimuthal modes depending on the magnetic field
orientation.

\section{Acknowledgements}
M.E. would like to thank B. Etemadi for useful discussions.

\bibliography{eem}

\newpage
\begin{center}{\bf Figure Captions}\end{center}
\noindent Figs. 1-3:  Bohmian trajectories corresponding to a
 two eigenfunctions superposition on an $R = 500\AA, a = 250
\AA$ torus with a circular cross section as a function of magnetic
field values (top to bottom)  $B_0 = 0, 2, 4$ Tesla.  \vskip 24 pt

\noindent Fig. 4: $\varepsilon_0$ for an $\alpha = .5, \beta = .1$
elliptical torus plotted as a function of magnetic flux $\gamma =
.263 B_0$. Diamonds correspond to the $U_C = 0$ case, stars to
$U_C \neq 0$, and squares to a two dimensional annular region with
inner radius $1-\alpha$ and outer radius $1+\alpha$. \vskip 24 pt

\noindent Fig. 5: $\varepsilon_0$ for an $\alpha = .1, \beta = .5$
elliptical torus plotted as a function of magnetic flux $\gamma =
.263 B_0$. Diamonds correspond to the $U_C = 0$ case, stars to
$U_C \neq 0$, and squares to a two dimensional vertical
strip/ribbon with unit radius and height $2\beta$.\vskip 24 pt

\noindent Fig. 6: $\varepsilon_0$ for an $\alpha = .5$ circular
torus plotted as a function of magnetic flux $\gamma = .263 B_0$.
Diamonds correspond to the $U_C = 0$ case, stars to $U_C \neq 0$.
\vskip 24 pt

\noindent Fig. 7: Detailed plot of $\varepsilon_0$ for an $\alpha
= .5, \beta = .1$ elliptical torus plotted as a function of
magnetic flux $\gamma = .263 B_0$. Stars correspond to the limit
of the two-dimensional annular region with inner and outer radii
$1\mp\alpha$ and diamonds to the elliptical torus with $U_C \neq
0$. \vskip 24 pt

\noindent Fig. 8: Detailed plot of $\varepsilon_0$ for an $\alpha
= .1, \beta = .5$ elliptical torus plotted as a function of
magnetic flux $\gamma = .263 B_0$. Stars correspond to the limit
of the two-dimensional vertical strip/ribbon of unit radius and
height $2\beta$, and diamonds to the elliptical torus with $U_C
\neq 0$. \vskip 24 pt

\newpage

\begin{figure}
\centering
\includegraphics{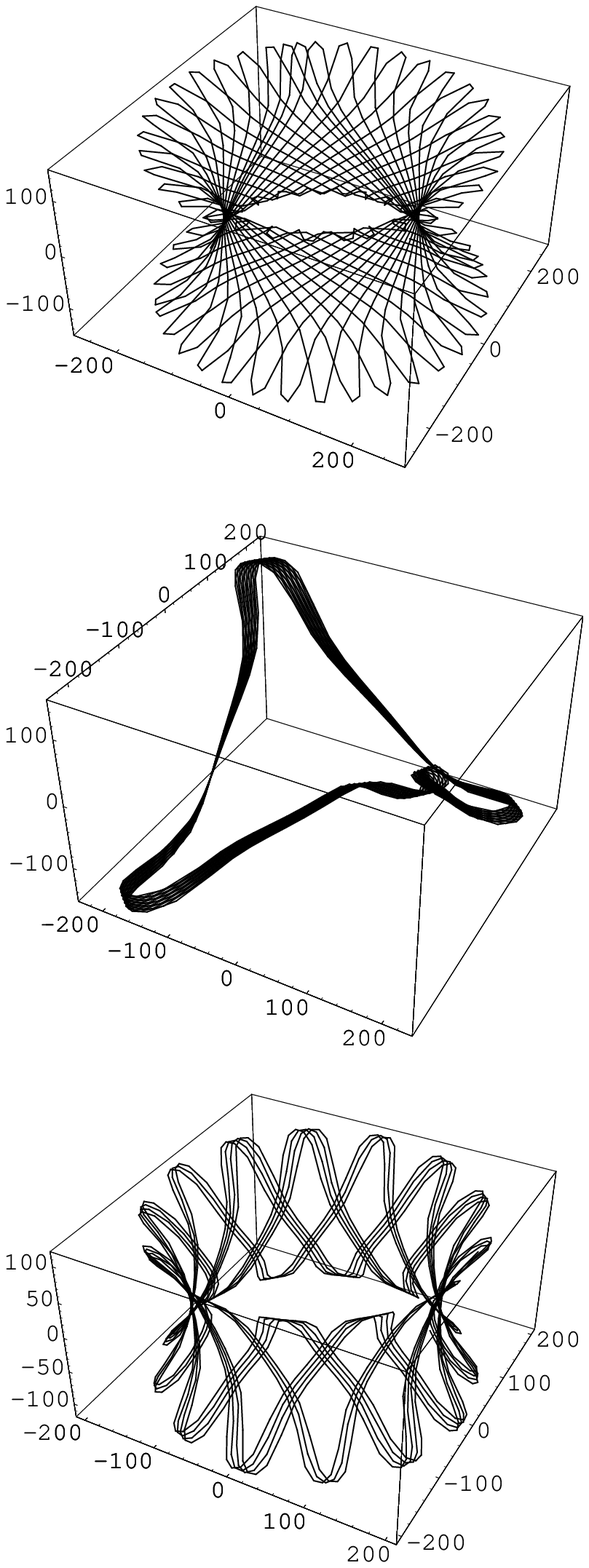}
\centerline{Figs. 1-3}
\end{figure}

\begin{figure}
\centering
\includegraphics{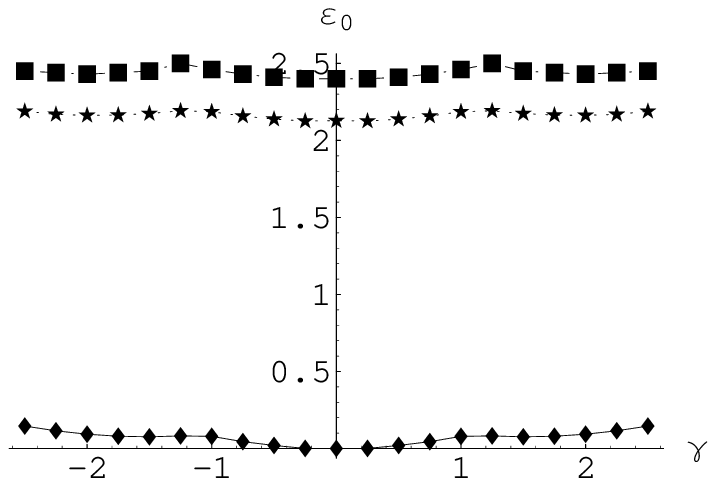}
\centerline{Fig. 4}
\end{figure}

\begin{figure}
\centering
\includegraphics{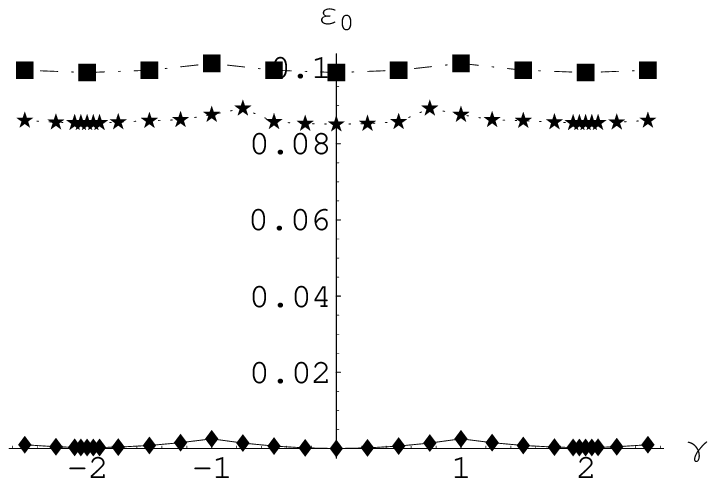}
\centerline{Fig. 5}
\end{figure}

\begin{figure}
\centering
\includegraphics{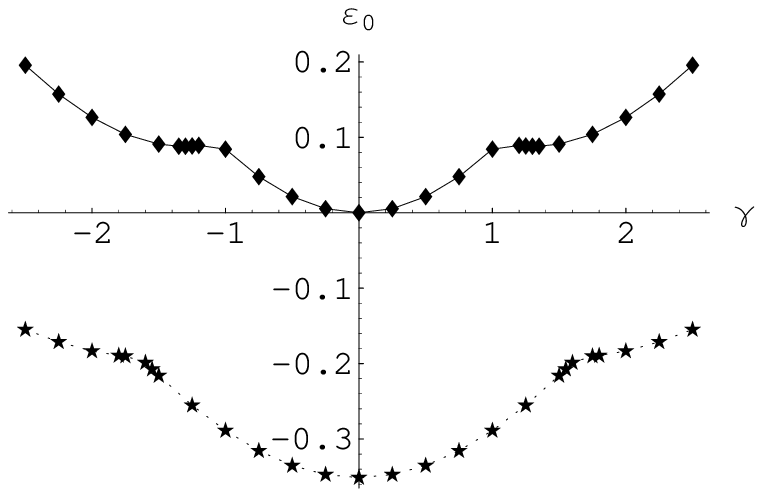}
\centerline{Fig. 6}
\end{figure}

\begin{figure}
\centering
\includegraphics{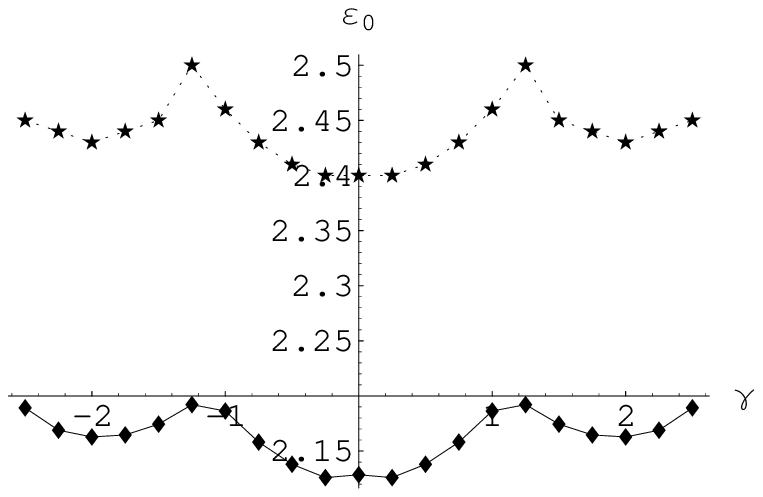}
\centerline{Fig. 7}
\end{figure}

\begin{figure}
\centering
\includegraphics{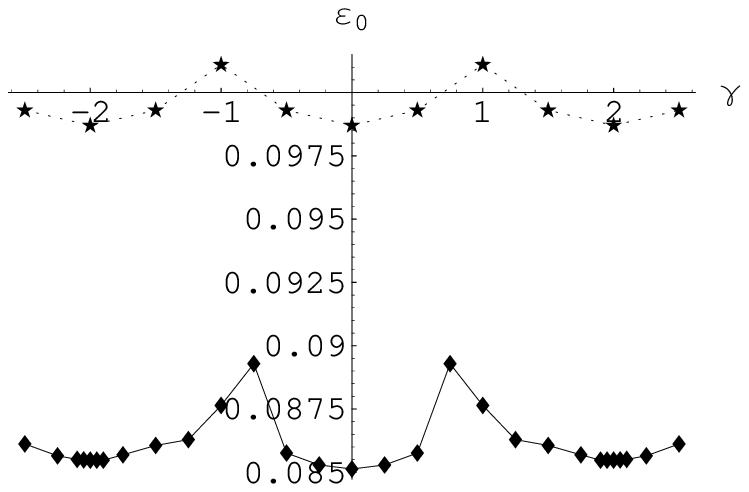}
\centerline{Fig. 8}
\end{figure}


\end{document}